\documentclass[reprint,aps,pre,floatfix,twocolumn,showpacs,superscriptaddress
]{revtex4-1}

\usepackage{comment}
\usepackage{type1cm}     
\usepackage{graphicx}     
\usepackage{xspace}     
\usepackage{balance}     
\usepackage{booktabs}     
\usepackage{multi row}     
\usepackage{bold-extra}     
\usepackage{siunitx}          
\usepackage[vlined,linesnumbered,ruled,noend]{algorithm2e}     
\usepackage{microtype}    
\usepackage{units}     
\usepackage{mathtools}     
\usepackage{amssymb}     

 \usepackage[hidelinks,unicode=true]{hyperref}
 \hypersetup{colorlinks=true,
 	linkcolor=blue,
 	urlcolor=blue,
 	citecolor=blue,
 	pdfhighlight=/N
 }

\usepackage[table,xcdraw]{xcolor}
\usepackage{subcaption}
\usepackage[font=small,labelfont=bf,
tableposition=top,
   justification=justified,
   format=plain]{caption}
 
\begin{document}

\title{Human Mobility in Response to COVID-19 in France, Italy and UK}

\author{Alessandro Galeazzi}
\email{a.galeazzi002@unibs.it}
\affiliation{University of Brescia, Via Branze, 59, 25123 Brescia , Italy}
\author{Matteo Cinelli}
\affiliation{Ca'Foscari Univerity of Venice, via Torino 155, 30172 Venezia, Italy}
\author{Giovanni Bonaccorsi}
\affiliation{Department of Management, Economics and Industrial Engineering, Politecnico di Milano}
\author{Francesco Pierri}
\affiliation{Department of Electronics, Information and Bioengineering, Politecnico di Milano}
\author{Ana Lucia Schmidt}
\affiliation{Ca'Foscari Univerity of Venice, via Torino 155, 30172 Venezia, Italy}

\author{Antonio Scala}
\email{antonio.scala.phys@gmail.com}
\affiliation{Applico Lab -- ISC CNR, Via dei Taurini 19, 00185 Roma, Italy}

\author{Fabio Pammolli}
\affiliation{Department of Management, Economics and Industrial Engineering, Politecnico di Milano}
\affiliation{CADS, Joint Center for Analysis, Decisions and Society, Human Technopole}

\author{Walter Quattrociocchi}
\email{w.quattrociocchi@unive.it}
\affiliation{Ca'Foscari Univerity of Venice, via Torino 155, 30172 Venezia, Italy}

\begin{abstract}
The policies implemented to hinder the COVID-19 outbreak represent one of the largest critical events in history. 
The understanding of this process is fundamental for crafting and tailoring post-disaster relief.
In this work we perform a massive data analysis, through geolocalized data from 13M Facebook users, on how such a stress affected mobility patterns in France, Italy and UK. 
We find that the general reduction of the overall efficiency in the network of movements is accompanied by geographical fragmentation with a massive reduction of long-range connections. The impact, however, differs among nations according to their initial mobility structure. 
Indeed, we find that the mobility network after the lockdown is more concentrated in the case of France and UK and more distributed in Italy. 
Such a process can be approximated through percolation to quantify the substantial impact of the lockdown. 
\end{abstract}

\maketitle

\section*{Introduction}
\label{sec:intro}

The pandemic outbreak of the COVID-19 virus has resulted in a unprecedented global health crisis with high fatality rates and heavy stress to national health systems~\cite{remuzzi2020covid,lai2020effect} and to the economic and social structure of countries~\cite{vespignani2020reshaping, colizza2020france}. 
As a result, an impressive effort is being exerted to understand on one side the epidemiological features of the outbreak~\cite{chinazzi2020effect,kraemer2020effect,oliver2020mobile,gatto2020spread} and on the other side its economic consequences~\cite{atkeson2020will,anderson2020will,mckibbin2020global}. 
The majority of countries governments have responded with non-pharmaceutical interventions (NPI) aimed at reducing the mobility of citizens to decrease the rate of contagion \cite{davies2020effect}. 
This calls for a better understanding of the patterns of human mobility during emergencies and in the immediate post-disaster relief.
Indeed the study of mobility habits is a foundational instance for several issues ranging from traffic forecasting, up to virus spreading and urban planning \cite{gonzalez2008understanding,dong2020statistical,eubank2004controlling,sciencemobility}. 
However, a quantitative assessment of its statistical properties at different geographical scales remains elusive \cite{brockmann2006scaling,havlin2002diffusion,song2010modelling,song2010limits,barbosa2018human}. 
The availability of rich datasets on mobility of individuals, coupled with the urgency of the current situation, has fostered the collaboration between tech giants, such as Facebook and Google, institutions and scholars \cite{oliver2020mobile,wellenius2020impacts,m2020democracy,bonaccorsi2020evidence}. 
Along this path, the present work builds upon a collection of data from social network users and addresses the dynamics of spatial redistribution of individuals as a response to mobility restrictions applied to limit the disease outbreak. 
We perform a massive analysis on aggregated and de-identified data provided by Facebook through its Disease Prevention movement maps \cite{maas2019facebook} to compare the effects of lockdown measures applied in France, Italy and UK in response to COVID-19. 
The overall dataset spans over 1 month of observations and accounts for movements of over 13M people. 
We model countries as networks of mobility flows and we find that restrictions elicit a transition toward local/short run connections, thus causing a loss in the network efficiency. 
Our result mirrors previous results in the literature which found that critical phenomena are an intrinsic feature of mobility networks, leading to transition between isolated short-range flows and collective long range flows~\cite{li2015percolation}.  Moreover we contribute to the literature on mobility disruption during critical events~\cite{bagrow2011collective,lu2012predictability,wang2014quantifying,wang2016patterns,martin2017leveraging} by studying the effect of movement limitations across the whole territory of the countries in our dataset, with a geographic scope which is unparalleled in the literature.  

We provide a model that simulates the effects of movement restrictions, finding that transitions can be approximated by means of different percolation strategies.
The general reduction of the overall efficiency in the network of movements is accompanied by geographical fragmentation with a massive reduction of long-range connections. 
However, the effect changes according to the starting mobility structure of each nation. 
In particular, we find that the network of UK and France after the lockdown is more concentrated while in Italy is more distributed. 
We conclude the paper by showing how the the effect of restrictions in Italy, UK and France can be approximated through percolation analysis.
Furthermore, our analysis reveals several interesting features. First, the three countries display differentiated mobility patterns that reflect the structural diversity in their underlying infrastructure: more centralized around their capital cities in the case of France and UK and more clustered in the case of Italy. 
Such infrastructural characteristics, together with different responses to national lockdown, contributed to the emergence of very distinctive configurations in terms of residual mobility patterns.
France has one big cluster centered in Paris and many smaller centers that disconnect as soon as the percolation process begins, Italy exhibits four interconnected clusters, centered approximately in Napoli, Roma, Milano and Torino, that remain interconnected over time thus showing a high persistence and resilience. 
Finally, UK has one cluster centered around London, but most of England exhibits a higher persistence with respect to France and Italy, thus suggesting the presence of a more capillary network structure. 

The understanding of the different resilience features of national mobility networks is fundamental to craft and tailor specific release policies and to smooth the economic impact of lockdown.
Indeed, the correlation among mobility, disease spreading and economic prosperity is crucial both in emergency scenarios and in ordinary times, since the different resilience of mobility networks could be both a predictor of the severity of future systemic crises and a guide to improve the economic and social impact of policies.

\section*{Connectivity of National Mobility Networks}
We represent national mobility networks as weighted directed graphs, based on movement maps made available by Facebook through their “Data for Good” program~\cite{maas2019facebook} (see Materials and Methods for further details). Nodes correspond to municipalities and edges are weighted with the amount of traffic between two municipalities. 

\begin{figure*}[h!]
\centering
\includegraphics[width=0.99\linewidth]{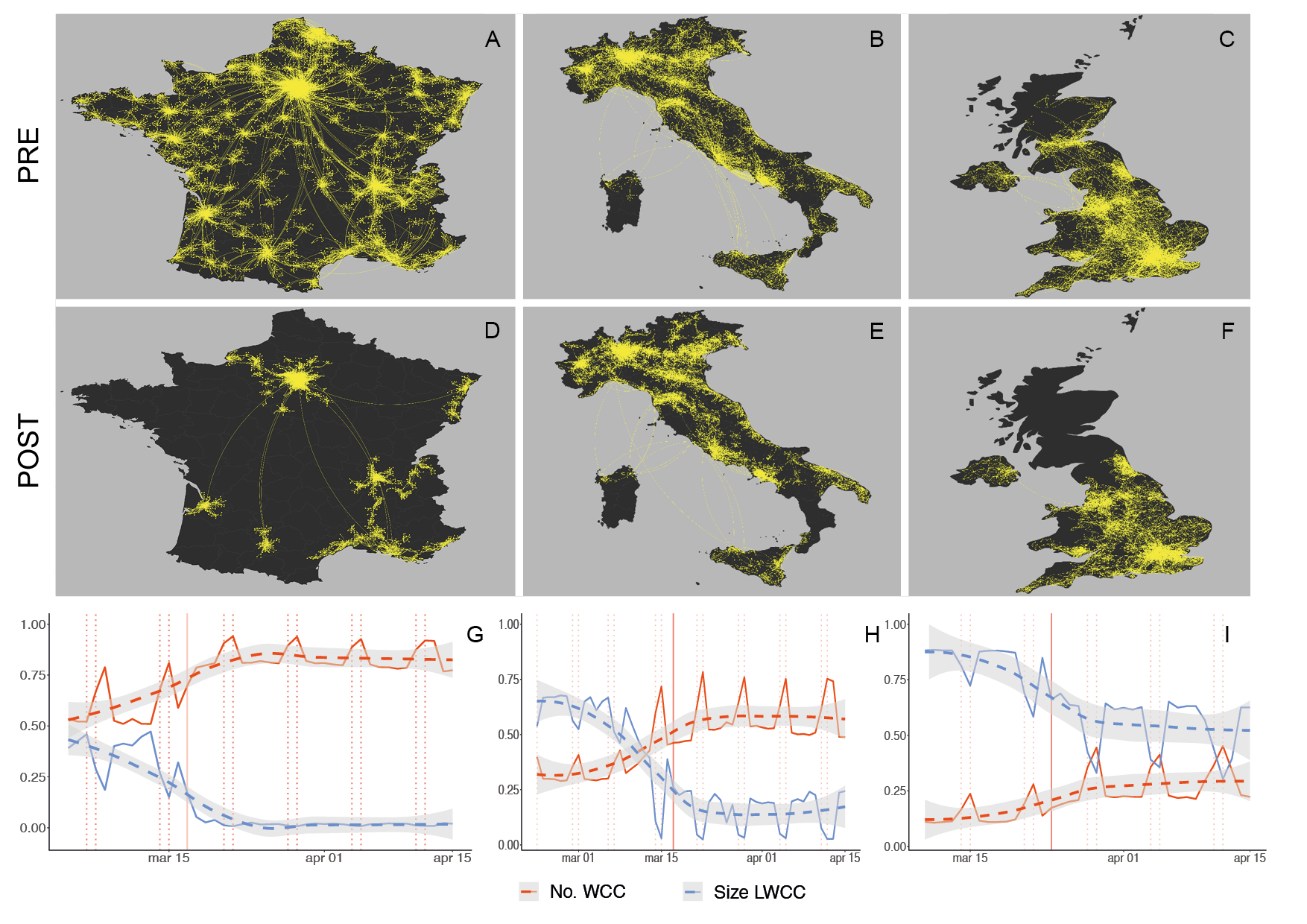}
\caption{\textbf{Outlook on national mobility networks for France, Italy and UK during COVID-19 pandemic.}\\
Panels (A to F) show the largest weakly connected components (LWCC) of national mobility networks built on two disjoint symmetric windows: respectively 2 weeks before (panels A-B-C) and 2 weeks after (panels D-E-F) the day of national lockdown. The lockdown dates is March 17th for France; March 9th for Italy; March 24th for UK. Bright dots represent municipalities that belong to the LWCC. We observe the following reductions in terms of nodes that disappear from the main cluster. France: from 5,495 to 1,174 nodes. Italy: from 2,733 to 2,293 nodes. UK: from 1,072 to 844 nodes. Panels (G-H-I) show the temporal evolution of daily connectivity for national mobility networks of municipalities, in terms of number of weakly connected components (No. WCC) and size of the LWCC. We visualize trends by means of a LOESS regression (dashed lines with 95\% confidence intervals shaded in grey) and highlight lockdown and week-end days with vertical red lines, respectively solid and dashed. 
}
\label{fig1}
\end{figure*}

We first aggregate mobility flows in two symmetric disjoint windows before and after the day of lockdown (see Materials and Methods), as shown in panels (A to F) of Figure~\ref{fig1}. By comparing the mobility network during the pre-lockdown phase (panels A-B-C of Figure~\ref{fig1}) to the mobility network during the post-lockdown phase, we note a significant reduction of the overall connectivity.
However, we notice that mobility restrictions have a higher impact on the connectivity of France, whereas they yields more limited effects in the other two countries. In fact, UK and Italy show a reduction of 21\% and 16\% in the size of the largest weakly connected component (LWCC, that is, the maximal subgraph of a network in which any two vertices are interconnected through an undirected path), whereas France exhibits a reduction of almost 79\%. 
Such different impact on the LWCC across countries may depend on several co-existing factors, among which we can list the structural features of the underlying networks and the population density, that is 199,82/km$^2$ for Italy, 270.7/km$^2$ for UK and 101/km$^2$ for France.
In panels G-H-I of Figure~\ref{fig1}, we further characterize daily connectivity patterns by computing the number of weakly connected components (No. WCC) and the size of the LWCC of the mobility networks. In all cases, we observe a decreasing trend on the size of LWCC and an increase of the number of WCCs. It is worth noticing that these trends are present even before the lockdown dates, but they reach a steady state only during the days after the intervention. 
Hence, the number of WCCs and the size of the LWCC  well capture the response of the countries to the lockdown. 
In the case of France (where the LWCC is Paris-centric), we notice a strong fragmentation of the network from the beginning: in fact, the number of WCCs is larger than the size of the LWCC, signaling that the mobility is well distributed along the whole country. However, Paris remains connected by long-range connections to the remaining most active areas of Bordeaux, Toulouse, Marseille and Lyon. In the case of Italy (where the LWCC contains all the main Italian cities, i.e. Napoli, Roma, Milano and Torino), the lockdown enhances the importance of local mobility.  Notice that, while in the centre and the south of Italy mobility remains localised mainly at regional level (in Fig. 1E one can notice Sicilia, Campania, Lazio and Toscana), the lockdown reveals that northern Italy is more interconnected, showing a clustered mobility for the main industrial regions, i.e. Piemonte, Lombardia, Veneto and Emilia-Romagna. The UK is clearly London-centric: the size of the LWCC remains higher than the number of WCCs, with strong local mobility patterns only in the Bristol area and in the Manchester-Liverpool area. Hence, the situation in terms of number of WCCs and the size of the LWCC reflects the different underlying structure of the three countries: France with a huge hub in Paris that is star-connected via long-range links to the local city-centered areas, Italy with mobility distributed mostly over the center-northern region, and UK that appears as an extension of London, whose mobility network remains pervasive even after the lockdown.

\section*{Efficiency of National Mobility Networks}

\begin{figure}[h!]
    \centering
    \includegraphics[scale=0.35, trim={2cm 1cm 0cm 1cm, clip}]{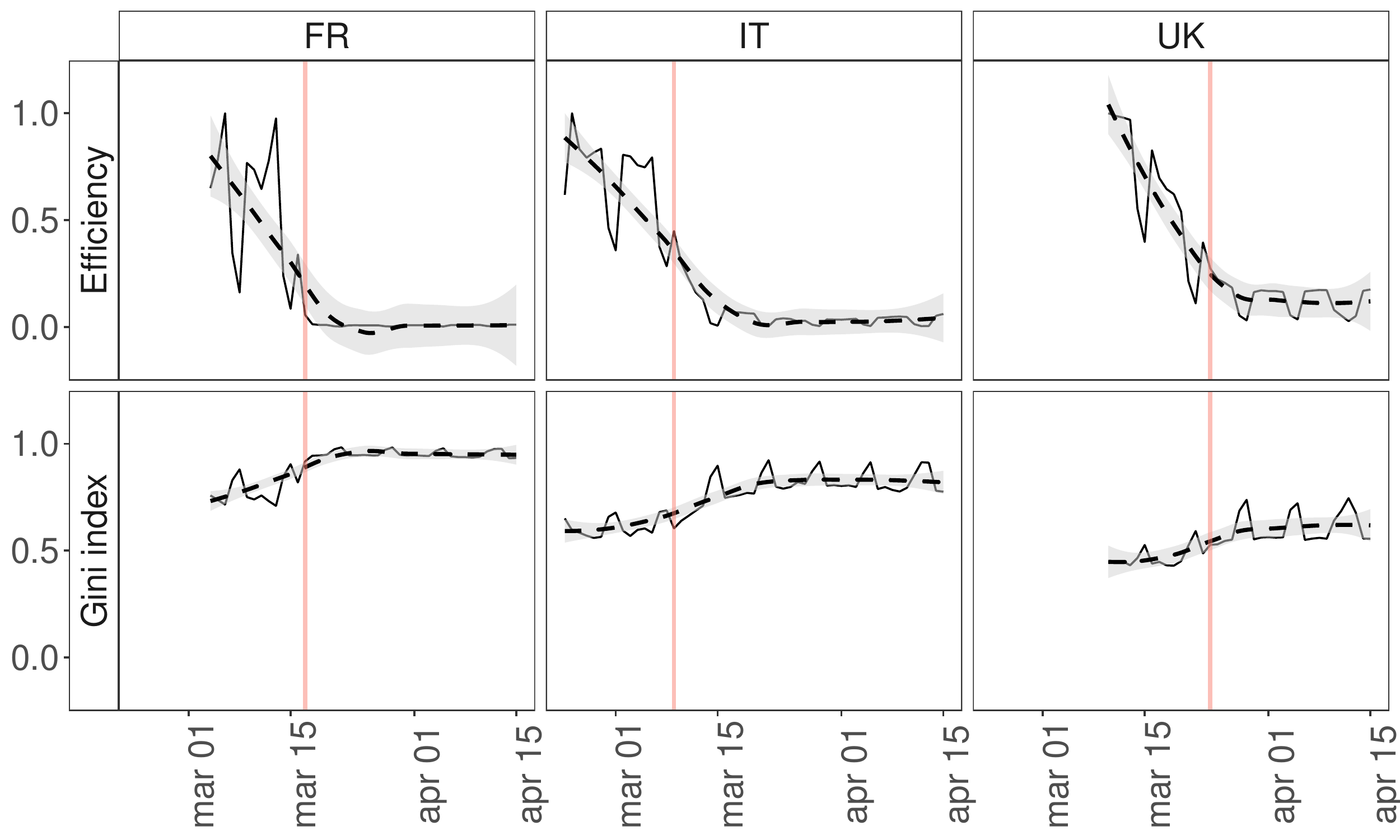}
    \caption{\textbf{Evolution of global efficiency and of its heterogeneity.}\\
    Top panels display the temporal evolution of global efficiency (normalized by its maximum value during the period of observation), for the mobility network of municipalities. 
    We visualize trends using a LOESS regression (dashed lines with 95\% confidence intervals shaded in grey) and highlight lockdown dates using a solid vertical line. 
    Bottom panels display the temporal evolution of the Gini index of the nodal efficiency.
    The Gini index is used as measure of heterogeneity and it is computed considering the nodal contributions to global network efficiency.
    Overall, we observe an increase of the Gini index indicating an increasing heterogeneity over time.}
    \label{fig:eff_gini}
\end{figure}

We further investigate the effect of lockdown focusing on the global efficiency~\cite{latora2001efficient} of mobility networks, as shown in Figure~\ref{fig:eff_gini}. The global efficiency is a measure that quantifies how optimal is the information flow in a network (further details are reported in Materials and Methods).

We notice a decreasing trend of the efficiency in the period before national lockdown and a steady state in the days after the intervention (Figure~\ref{fig:eff_gini}), which is consistent with the observed decrease in global connectivity (Figure~\ref{fig1}).
 
Since the network efficiency is a measure that condensates information related to both clustering (i.e. connectedness of neighbours) and small world effect (i.e. presence of long-range connections that act as shortcuts), the trends observed in  Figure~\ref{fig:eff_gini} (top panels) well describe the effect of an ongoing shock that cuts both long range connections and the overall cohesiveness.
Moreover, the three countries experience different amounts of decentralization as a consequence of the lockdown.
In order to capture the differences among them in more detail, we measure the heterogeneity of nodes in terms of their contribution to the global efficiency using the Gini index~\cite{gini1921measurement} (see Materials and Methods).
We first observe that, in correspondence of the initial decrease of global efficiency, the Gini index displays an increasing trend that becomes steady after the lockdown date (Figure~\ref{fig:eff_gini}). The observed trend means that the contribution of nodes to the global efficiency becomes more and more heterogeneous over time, until it reaches a steady state. This result indicates that the progressive disruption has very heterogeneous effects on the single nodes, suggesting that policies should be carefully tailored to avoid enhancing unequal treatments of different areas: as an example, in Italy it has been observed that the lockdown could enhance economic disparities~\cite{bonaccorsi2020evidence}. Such an effect could be even more pronounced in France, that shows a higher level of dishomogeneity in connectivity both at the municipal and at the provincial level when compared to Italy.
Indeed, nodes that get disconnected are more likely to remain isolated as compared to a country such as Italy that is based on a more distributed mobility network
Conversely, UK seems to have a different response to the shock distributing it more evenly among its nodes (refer to SI for further details).

\section*{Resilience of National Mobility Networks}
The drivers behind the empirical results reported in previous sections can be investigated by modeling the process that led to the disruption of mobility networks. Therefore, in order to model the lockdown effect, we perform an analysis based on percolation theory \cite{essam1980percolation} on the aggregated graph related to the period before the national lockdown. We assume that such a graph is a proxy for the structure of the mobility network in standard conditions (see Materials and Methods).
We implement bond percolation on the aggregated networks by iteratively deleting edges following an increasing (respectively decreasing) weight order. 
During the process of network dismantling, we keep track of measures related to both cohesiveness and distance, namely the LWCC size, the global efficiency and the node persistence. In more detail, the node persistence measures the extent to which a node remains connected to the LWCC, that is, how much the node is resilient to percolation (see Materials and Methods).

\begin{figure*}[h!]
    \centering
    \includegraphics[width=0.99\linewidth]{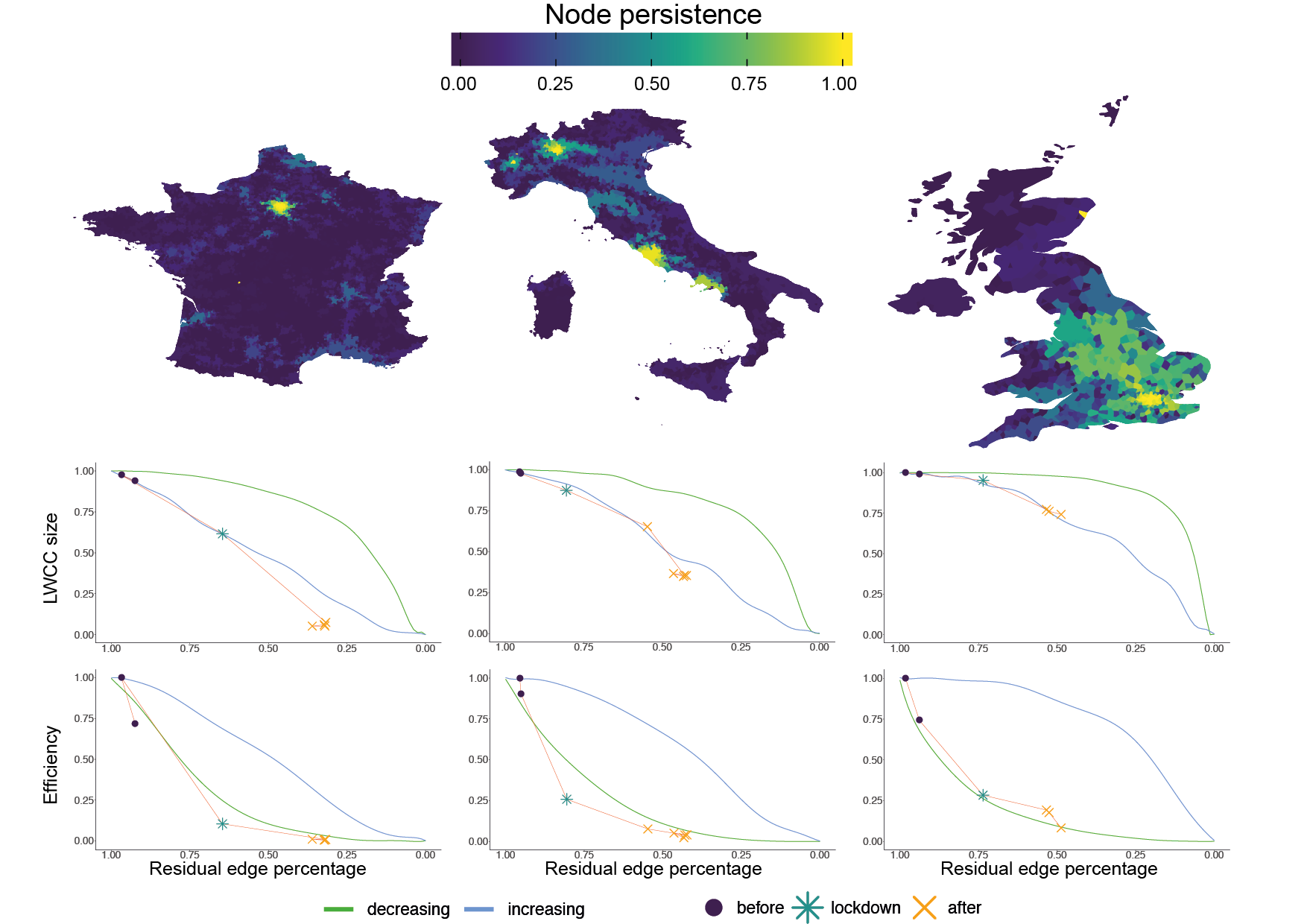}
    \caption{\textbf{Results of the percolation process in terms of node persistence and edge removal strategies }.\\
    Percolation process is performed by iterating a cutting procedure on edges according to their weights calculated in the whole period before lockdown.
    Top panels: node persistence during the percolation process. Node persistence is defined as the number of iterations a node remains connected to the LWCC over the maximum number of iterations before the network is completely disconnected; the more persistent the node, the brighter the color. Notice that we are defining persistence upon deleting edges by their increasing strength: hence, brighter nodes are not only the last to be disconnected, but are also those embedded in stronger mobility flows. 
    Central panels: size of the LWCC (largest weakly connected component) as a function of the residual edges. Green curves correspond to deleting edges from the strongest to the weakest, while blue curves correspond to deleting edges according to their increasing weights. 
    Bottom panels: variation of global network efficiency obtained deleting edges by increasing (green curves) or decreasing (blue curves) strength.
    In both the central and the bottom panels, we plot the "empirical" values of LWCC and of the global efficiency for the mobility networks calculated aggregating flows in the weeks before (circles), during (triangles) and after (diamonds) the lockdown date. Notice that, while lockdown has cut out the peripheries of the networks (central panels, but see also panels D-E-F of Figure~\ref{fig1}), the effect of the reduced mobility also severely affects the network efficiency (lower panels).
    }
    \label{fig:voronoi}
\end{figure*}
The top row of Figure~\ref{fig:voronoi} shows the results of the percolation process in terms of node persistence, carried out by removing edges in increasing weight order for France, Italy and UK. To each node we assign an area on the map calculated by means of Voronoi tessellation~\cite{voronoi1908nouvelles} and colored according to node persistence (see Materials and Methods). 




The empirical evidence displayed in previous sections finds support in the percolation results. Comparing the three cases, the differences in the network structure among countries clearly emerge: while France has one big cluster centered in Paris and many smaller centers that disconnect as soon as the process proceeds, Italy exhibits four interconnected clusters, centered approximately in Napoli, Roma, Milano and Torino (that roughly correspond to the high speed train lines), that remain interconnected thus showing a high persistence. Conversely, UK has one cluster centered around London, but most of England exhibits a higher persistence with respect to France and Italy, thus suggesting the presence of a more capillary network structure. 
Surprisingly, deleting stronger edges first does not disconnect any of the three networks as fast as deleting weaker edges. This effect can be explained by assuming a "rich club" structure, where links corresponding to higher mobility flows are concentrated around core regions.
Indeed, the largest fragility of the mobility network toward the deletion of weaker edges hints at a core-periphery structure.

To provide further insights regarding the different effects of the percolation process on each of the three countries, in the middle row Figure~\ref{fig:voronoi} we compare the trends of LWCC size and global efficiency throughout the percolation process. 

We first notice that the decay of the LWCC size differs depending both on country and edge removal strategy: removing edges sorted by decreasing weight seems to affect less the decay of the LWCC size than removing edges sorted by increasing weight, for all countries. However, we notice some differences in the decay of LWCC size among the three nations in the increasing case: while France exhibits an almost linear trend, UK shows a significant drop only after a notable amount of connections is removed. Finally, Italy seems to be in a intermediate condition, showing an almost linear decay interspersed with steps.




The bottom row of Figure~\ref{fig:voronoi} displays the normalized global efficiency as a function of the residual edge percentage. Also in this case, France has a smoother decay with respect to Italy and UK, especially for what concerns the percolation process based on increasing weight sorting. Moreover, fixed the percentage of residual edges, UK has a higher efficiency when we consider the increasing percolation case. However, such a relationship changes when the decreasing percolation strategy is taken into account: in this case UK has a somewhat steeper decay of global efficiency, suggesting the presence of a higher number of high weight connections among nodes. 
Again, we observe that core regions (the ones where strong links are concentrated, i.e. the most persistent regions in the upper panels) are the most relevant: in fact, the steepest decrease in efficiency happens when deleting strongest edges first. On the other hand, periphery regions only mildly contribute to the network efficiency, as demonstrated from the slowest decrease upon cutting weakest edges first.

In order to shed light on the impact of the lockdown on the network edges, we report the trends followed by empirical networks superimposing them over the results of the percolation process.
Each point in the curve corresponds to a certain statistic computed on a mobility network aggregated over one week (see Materials and Methods). The mobility networks are divided into three categories based on the period taken into account: the weeks before the lockdown, the week in which the lockdown happened and the weeks after the lockdown.

We note that the residual amount of edges decreases over time and it differs from country to country. France displays the lowest percentage of residual edges, while Italy and UK tend to retain a higher percentage of their edges.
Moreover, we notice how Italy and UK strongly differ in terms of the evolution of LWCC size: although they loose roughly the same fraction of edges, there is a strong difference in terms of the amount of connected nodes. In spite of such differences, the increasing percolation strategy is able to approximate the reduction in terms LWCC size quite accurately for all nations.
The case of the normalized global efficiency is somewhat different: all the countries seems to have a similar loss proportionally to their initial efficiency. Additionally, the trend of the empirical curves is closer to that of the decreasing percolation strategy.
The major difference observed in the middle and bottom rows of Figure~\ref{fig:voronoi} is based on the fact that the considered empirical statistics follow opposite link removal strategies. This effect is due in part to the nature of the two measures; the former more related to the density and to the structure of connections and the second more related to their weight. 
In any case, despite eventual differences related to the network measures that we take into into account, it is interesting to observe how the lockdown cannot be modeled by a removal strategy based only on edge weights. Rather, it is the result of a joint effect deriving from the removal of links based both on their structural importance and weight.




\section*{Conclusions}
The COVID-19 pandemic is testing the structural strength of our global society. Most national governments have simultaneously reacted to the contagion by applying mobility restrictions to contain the disease outbreak. The resulting disruption is similar to that caused by a natural disaster, but the effect is on a global scale. For this reason, in this work we analyze the effects of the lockdown on three different national mobility systems, the French, Italian and British, by means of a 13M users dataset from Facebook. We provide a structural analysis of the mobility network for each nation and quantify the effect of mobility restrictions applied to hinder COVID-19 outbreak by means of percolation analysis. We find that lockdown mainly affects national smallwordness – i.e., strong reduction of long-range connections in favor of local paths.  Our analysis suggests that the national resilience to massive stress differs and depends upon the inner connectivity structure. 
Indeed, the three countries display very different mobility patterns that reflect the diversity in their underlying infrastructure, more concentrated in the case of France and UK and more distributed in the case of Italy.
Our results may provide insights on the interplay between mobility patterns and structural dependencies in the network of movements at national level and may help design better transportation networks, improve system efficiency during natural catastrophes and contain epidemics spreading.

\section*{Materials and Methods}
\label{sec:mm}
\small{
\subsection*{Mobility data and networks} We analyzed human mobility leveraging data provided by Facebook through its “Data for Good” program~\cite{maas2019facebook}. The platform provides movement maps which are based on de-identified and aggregated information of Facebook users who enabled their geo-positioning. 

Movements across administrative regions (i.e. municipalities in our case) are aggregated with a 8-hours frequency, and describe the amount of traffic flowing between two municipalities in a given time window. Similar to data analyzed in recent research on mobility restrictions applied in China~\cite{kraemer2020effect,chinazzi2020effect}, Facebook does not really provide the number of people moving between two locations but rather an index, constructed with proprietary method to ensure privacy and anonymization, that highly correlates with real movements~\cite{maas2019facebook}. 

We collected data relative to mobility in Italy, France and United Kingdom until April 15th, with different starting times depending on the availability of Facebook maps (respectively February 23th, March 10th and March 4th). The average number of daily users with their location enabled during the period of interest is 13,669,145 (France: 4,110,226; UK: 5,801,979; Italy: 3,786,940).

For the sake of our analysis we represent mobility flows using a weighted directed graph where nodes are municipalities and edges are weighted based on the amount of traffic flowing between two locations. To represent mobility networks before/during lockdown we aggregate daily traffic on windows of 12-13-14 days (respectively for France, UK and Italy) before/after the day of intervention depending on the availability of data.

From these data, we build several graphs for each country. We consider an oriented daily graph at municipality levels, where there can be only one edge per day from municipality A to municipality B, whose weight is the daily sum of all edges from A to B. 
The same procedure is applied for building aggregate networks at different time scales such as weekly graphs of pre/post lockdown graphs.

\subsection*{Network efficiency}
The efficiency is a global network measure that combines the information deriving from the network cohesiveness and the distance among the nodes.
It measures how efficiently information is exchanged over the network \cite{latora2001efficient} and it can be defined as the average of nodal efficiencies $e_{ij}$ among couples of vertices of the network. Given a weighted network $G(V,E)$ with $n=|V|$ nodes and $m=|E|$ edges, the connections of $G$ are represented by the weighted adjacency matrix $W$ with elements $\{w_{ij}\}$ where $w_{ij} \geq 0$ $\forall$ $i,j$.
The global efficiency can be written by means of the following expression:
\begin{equation}
    E_{glob}(G) = \frac{1}{n(n-1)}\sum_{i \neq j \in V}e_{ij} = \frac{1}{n(n-1)}\sum_{i \neq j \in G} \frac{1}{d_{ij}}
\end{equation}
where $d_{ij}$ is the distance between two generic nodes $i$ and $j$, defined as the length of the shortest path among such nodes. The shortest path length $d_{ij}$ is the smallest sum of the weights $w_{ij}$ throughout all the possible paths in the network from $i$ to $j$. When node $i$ and $j$ cannot be connected by any path then $d_{ij} = +\infty$ and $e_{ij} = 0$.
Following the methodology of~\cite{latora2001efficient}, the global efficiency $E_{glob}(G)$ is normalized in order to assume maximum value $E(G)=1$ in the case of perfect efficiency. In such a setting the nodal efficiency, i.e. the contribution of each node to the global efficiency, can be simply written as: 
\begin{equation}
\label{eqn:eglob}
    e_{i} = \frac{1}{n-1}\sum_{j \neq i}\frac{1}{d_{ij}} \,.
\end{equation}
Beside the geographical distance between two nodes of the graphs, a proximity can also be defined considering that two locations are closer if many movements happen between them. 
To compute network efficiency in our case we use the reciprocal of weights on links to obtain the shortest path distance among couples of nodes.

\subsection*{Gini index}

The Gini index is a classic example of a synthetic indicator used for measuring inequality of social and economic conditions.
The Gini index can be defined starting from the Gini absolute mean difference $\Delta$~\cite{kendall1958theadvanced} of a generic vector $y$ with $n$ elements, that can be written as:
\begin{equation}
\Delta = \frac{1}{n^2}\sum_{i=1}^n \sum_{j=1}^n |y_{i} - y_{j}|
\label{eq:delta}
\end{equation}
The relative mean difference is consequently defined as $\Delta/\mu_y$ where $\mu_y= n^{-1}\sum_{i=1}^n y_i$. Thus, the relative mean difference equals the absolute mean difference divided by the mean of the vector $y$. The Gini index $g$ is one-half of the Gini relative mean difference.
\begin{equation}
g = \frac{\Delta}{2 \mu_y}
\label{eq:Gini}
\end{equation}
Values of $g \sim 1$ signal that the considered vector displays high inequality in the distribution of its entries, while values of $g \sim 0$ signal a tendency towards equality.

\subsection*{Percolation process and node persistence}
To measure the extent to which a node resists to percolation process, we define the following quantity as node persistence. Consider a graph $G$ with nodes  $V$=$\{v_1,\dots,v_k\}$  and edges $E$=$\{e_1,\dots,e_h\}$. 
Suppose that the edges weights assume values in the set $W=[w_1,\dots,w_{n+1}]$ with $w_1<w_2<\dots<w_{n+1}$, and suppose that we run a percolation process that consists of deleting edges in increasing order, that is, in the first iteration we delete all the edges with weight less or equal than $w_1$, in the second iteration we delete all the edges with weights less or equal than $w_2$ and so on until the last iteration, $n+1$, where we delete all the edges in the network.
Notice that the percolation process can be performed similarly in the case edges are removed following a decreasing weight order.
Thus, at iteration $i$, we delete all the edges with weight less or equal than $w_i$. Consider now a node $v_j$ $\in$ $V$. At each step of the process, $v_j$ may or may not be part of the LWCC of the network $G$. Clearly, if $v_j$ does not belong to the LWCC of $G$ at step $i$, it will not be in the LWCC of $G$ at step $i+1,i+2,\dots,n+1$. Defining $M_{v_j}$ as the maximum number of iterations $i$ such that $v_j$ belongs to the LWCC of $G$, the persistence of node $v_j$ is defined as:
\begin{equation}
  \rho=\frac{M_{v_j}}{n}
\end{equation}

That is, if the whole process takes $n+1$ iterations to disconnect the whole network, the persistence of a node $v_j$ is defined as the maximum number of iterations for which $v_j$ is connected to the LWCC over the maximum number of iteration for which there are connected nodes in the network.

\bibliographystyle{unsrt}

\end{document}